\pdfoutput=1

\documentclass[final,5p,twocolumn]{elsarticle}

\usepackage{amssymb}
\usepackage{hyperref}

\biboptions{sort&compress}

\begin{document}

\begin{frontmatter}

\title{Collective flow in ultrarelativistic $^3$He - Au collisions}

\author[agh,ifj]{Piotr Bo\.zek}
\ead{Piotr.Bozek@ifj.edu.pl}

\author[ujk,ifj]{Wojciech Broniowski}
\ead{Wojciech.Broniowski@ifj.edu.pl}

\address[agh]{AGH University of Science and Technology, Faculty of Physics and Applied Computer Science, al. Mickiewicza 30, 30-059 Krakow, Poland}
\address[ifj]{The H. Niewodnicza\'nski Institute of Nuclear Physics PAN, 31-342 Krak\'ow, Poland}
\address[ujk]{Institute of Physics, Jan Kochanowski University, 25-406 Kielce, Poland}

\begin{abstract}
The  triangular flow in ultrarelativistic
$^3$He-Au collisions at RHIC energies is enhanced due to the triangular arrangement of the nucleon configurations in 
$^3$He.  We study the fireball eccentricities in the Glauber Monte Carlo model and find that
since the configurations of the projectile $^3$He are elongated triangles, the created fireball has a large ellipticity and a smaller triangularity. 
The  dependence of the triangularity on centrality is weak, so it cannot be extracted from the centrality dependence of the triangular flow $v_3$, 
as it is dominated by the centrality dependence of the hydrodynamic response. 
We propose to look at the centrality dependence of the ratio
 $v_n\{4\}/v_n\{2\}$, where the uncertainties from the
hydrodynamic response cancel, and show that the basic signature of the geometry-driven collective flow is the raise of the ratio  
$v_3\{4\}/v_3\{2\}$ with the number of participant nucleons for centralities less than 10\%. 
\end{abstract}


\end{frontmatter}

\section{Introduction}

Collective behavior in relativistic collisions of small system is an active field of experimental
studies at RHIC and the LHC   \cite{CMS:2012qk,Abelev:2012ola,Aad:2012gla,Adare:2013piz}
A large number of measurements are consistent with calculations in the hydrodynamic model 
 \cite{Bozek:2011if,Qin:2013bha,Bozek:2013uha,Werner:2013ipa,Nagle:2013lja,Kozlov:2014fqa}. 
Some observations can also be explained in the color class condensate framework~\cite{Dusling:2013oia,Dusling:2012wy}.
On-going studies are aimed at elucidating the nature of the observed flow correlations and 
test the limits of collectivity in small systems.

The azimuthal deformation of the fireball in small systems is due to fluctuations, as in p-Pb collisions, 
or to a combination of fluctuations and the intrinsic deformation of the small projectile, as in d-Au collisions.
Collisions involving a projectile with a triangular deformation, $^3$He-Au \cite{Sickles:2013mua} or $^{12}$C-Au 
\cite{Broniowski:2013dia} are particularly interesting, as they provide systems with a geometry-driven triangular flow.
The difficulty in the study of the geometry-driven flow in small systems comes from the interplay of a large contribution from the shape fluctuations
to the initial eccentricities of the fireball. While the large quadrupole deformation of the deuteron makes it possible to trigger on
central events to  get a sample of  events with a  large eccentricity~\cite{Bozek:2011if}, for $^3$He-Au collisions the  
centrality dependence of the triangularity is weaker and it is much more difficult to identify the triangular flow  driven by 
the projectile geometry~\cite{Nagle:2013lja}.

We study the eccentricities of the fireball formed in $^3$He-Au collisions as a function of centrality (here defined via the number of 
wounded nucleons~\cite{Bialas:1976ed}) to find  signatures of the 
triangular flow caused by the geometrical deformation the projectile.
We find that the effect is clearly seen in the ratio of the cumulant moments of the eccentricities,
$\epsilon_3\{4\}/\epsilon_3\{2\}$, thus suggesting to investigate the 
ratio $v_n\{4\}/v_n\{2\}$ in experimental studies. We show that the basic signature of the geometry-driven triangular flow is the raise of this ratio  
with the number of wounded nucleons for centralities below 10\%.

\section{Method \label{sec:method}}

The Fourier coefficients $v_n$ of the azimuthal dependence 
\begin{equation}
\frac{dN}{ d\phi} = \frac{N}{ 2 \pi}\left[ 1+2\sum_n v_n \cos\left(n(\phi-\Phi_n)\right)\right]
\label{eq:vn}
\end{equation}
of the spectra of particles emitted in relativistic nuclear collisions appear due to the collective expansion of
 an  azimuthally deformed  source profile  (in the following we consider the flow coefficients integrated over the transverse momentum).
The hydrodynamic evolution that generates the azimuthally asymmetric particle distribution gives an 
 approximately linear response of the flow coefficients $v_n$ to the eccentricities of the initial source density $\rho(x,y)$ in the transverse plane,
\begin{eqnarray} 
\epsilon_n e^{i n \Phi_n} = - \frac{\int \rho(x,y) e^{i n\phi}  (x^2+y^2)^{n/2} dx dy}{\int \rho(x,y) (x^2+y^2)^{n/2}  dx dy}, \label{eq:eps}
\end{eqnarray}
for $n=2,3$ \cite{Gardim:2011xv,Niemi:2012aj,Teaney:2012ke}, with $\phi=\arctan(y/x)$ and $\Phi_n$ denoting the angles of the principal axes.

The flow fluctuates from event to event.
The cumulant method allows one to extract even cumulant moments $v_n\{m\}$ of the distribution of flow coefficients 
$v_n$ \cite{Borghini:2000sa}. 
%
%
With the linear hydrodynamic response one has the proportionality
\begin{equation}
v_n = \kappa_n \epsilon_n \ , 
\label{eq:linear}
\end{equation} 
hence the cumulant flow coefficients can be related  to the corresponding moments of the eccentricity distributions in the initial state, namely
\begin{eqnarray}
v_n\{m\}=\kappa_n \epsilon_n\{m\},
\end{eqnarray}
where the response coefficient $\kappa_n$ is independent of the rank $m$, but it does depend of the dynamic features 
such as the multiplicity or the collision energy. 
We will need explicitly
\begin{eqnarray}
     \epsilon_n^2\{2\} &=& \langle \epsilon_2^2 \rangle, \nonumber \\
     \epsilon_n^4\{4\} &=& 2 \langle \epsilon_n^2 \rangle^2 -
\langle \epsilon_n^4 \rangle.
\end{eqnarray}

Relation~(\ref{eq:linear}) allows one to discuss the cumulant moments of the eccentricity 
instead of the flow coefficients, i.e., the features of the initial state can be used
to make certain predictions for the final flow coefficients. 
In particular, with the Glauber model of the initial state one finds a large ellipticity  $\epsilon_2$  for collisions with the deuteron projectile 
\cite{Bozek:2011if}, and a substantial triangularity  $\epsilon_3$ for collisions  with the $^3$He  \cite{Sickles:2013mua} or $^{12}$C  
\cite{Broniowski:2013dia} projectiles. The geometric deformation increases for collisions with a larger number 
of participants, corresponding to high multiplicity events. On the other hand, the 
eccentricity due to fluctuations of independent sources 
decreases with the number of participants.
We recall that for a finite number of wounded nucleons $N_{w}$  the eccentricity distribution is not of a Bessel-Gaussian \cite{Bhalerao:2006tp,Alver:2008zz}. In particular, $\epsilon_n\{m\}\neq 0$ for $m \ge 4$, 
and $\epsilon_n\{m\}$ decreases as $1/N_{w}^{1-1/m}$.
Accordingly, for p-Pb collisions a nonzero value of the higher order cumulants is expected from fluctuations 
\cite{Bzdak:2013rya,Yan:2013laa,CMS:2014bza}, which does not signal by itself an intrinsic geometric deformation of the source.

For events with a large number of participants, the contribution from fluctuations to $\epsilon_n$ decreases, while the geometrical deformation is enhanced 
due to the preferential orientation of the deformed projectile hitting the large nucleus~\cite{Broniowski:2013dia}.
This brings the possibility to identify the geometric
deformation in the initial state through the increase of $v_2$ or $v_3$ for the high-multiplicity events. 
Unfortunately, the argument cannot be applied directly, since the hydrodynamic response (\ref{eq:linear}) depends 
on the centrality, i.e., $\kappa_n$ increases with the multiplicity of the event. Therefore, just from the increase of $v_n$ with 
centrality one cannot infer that the deformation of the fireball grows as well.  
This is especially difficult for $^3$He-Au collisions, where, as we shall see, the increase of $\epsilon_3$ for central events is very mild. 

One possibility, of course, is to run the involved hydrodynamic simulations, as in \cite{Nagle:2013lja}. However, such modeling 
introduces the uncertainties of hydrodynamics, which for small systems may lead to substantial sensitivity and, in fact, difficulty 
in pinpointing the signatures of the geometric deformation of the initial state. 
We thus propose a different strategy to evidence the presence of an initial intrinsic deformation.
By considering the ratio of cumulants of different order 
for a given flow coefficient $v_n$, with $m \ge 4$, we
gain two things. First, the hydrodynamic response with unknown centrality dependence cancels out in the ratio
\begin{eqnarray}
\frac{v_n\{m\}}{v_n\{2\}}=
\frac{\epsilon_n\{m\}}{\epsilon_n\{2\}} \ , \label{eq:ratios} 
\end{eqnarray}
and the centrality dependence of the ratio of flow cumulant can be directly compared to the corresponding ratio of eccentricity cumulants.
Second, the ratio $\epsilon_n\{m\}/\epsilon_n\{2\}$ has a known behavior as a function of the number of participants in two important limits. 
For a fireball, with deformations solely driven  by fluctuation of independent sources, the ratio monotonously decreases 
as $N_{w}^{1/m-1/2}$, 
whereas if the fireball possesses an intrinsic geometric deformation, the ratio approaches
 $1$ from below for (very) large $N_{w}$.

\section{$^3$He wave functions and eccentricities \label{sec:he3}}

To obtain a large triangular deformation in $^3$He-Au collisions, two condition must be met. First,  the plane of the $^3$He 
nucleus should be more-less aligned with the transverse plane (flat-on collision), second, the configuration of the $^3$He wave-function should 
have a large triangularity, which happens for configurations close to an equilateral triangle. In practice, it is difficult to 
realize these conditions in a typical  event, which makes the experimental observation of the geometrical triangularity challenging.
Thus our first goal is to understand the structure of $^3$He in simple,
geometric terms. Similarly to Ref.~\cite{Nagle:2013lja}, we use the samplings of the
$^3$He wave functions
as provided, e.g., in the distribution of the Phobos Monte Carlo
code~\cite{Loizides:2014vua}, generated within the state-of-the-art Monte Carlo Green's function method
\cite{Carlson:1997qn}. 
We start our analysis with a closer look at these distributions. The centers of
the three nucleons form a triangle. We consider eccentricities defined 
by these three points, evaluated in the plane determined by the triangle.

\begin{figure}[tb]
\includegraphics[angle=0,width=0.33 \textwidth]{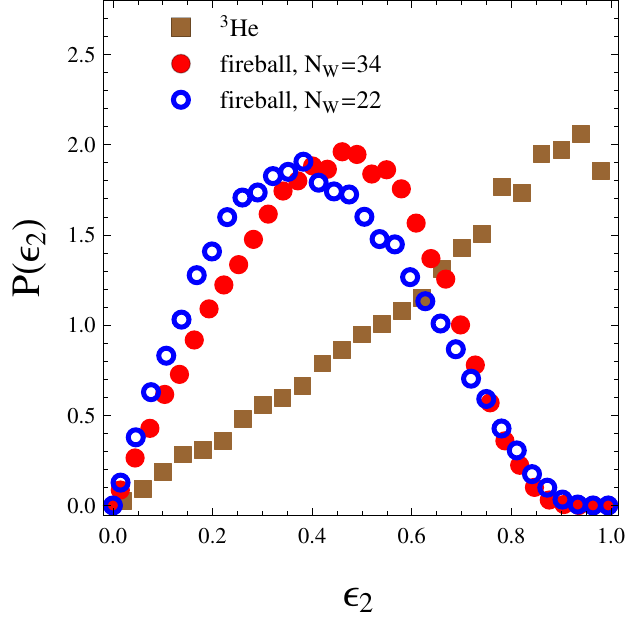}
\caption{Probability distributions of ellipticity of the nucleon distributions in $^3$He in the plane containing the three nucleons (squares), and in the fireball created in the collision with $^{197}$Au
at $N_{w}=22$  (open circles) and $N_{part}=34$  (filled circles) (wounded nucleon model, source smearing parameter 0.4~fm).
\label{fig:eps2}} 
\end{figure}  
\begin{figure}[tb]
\includegraphics[angle=0,width=0.33 \textwidth]{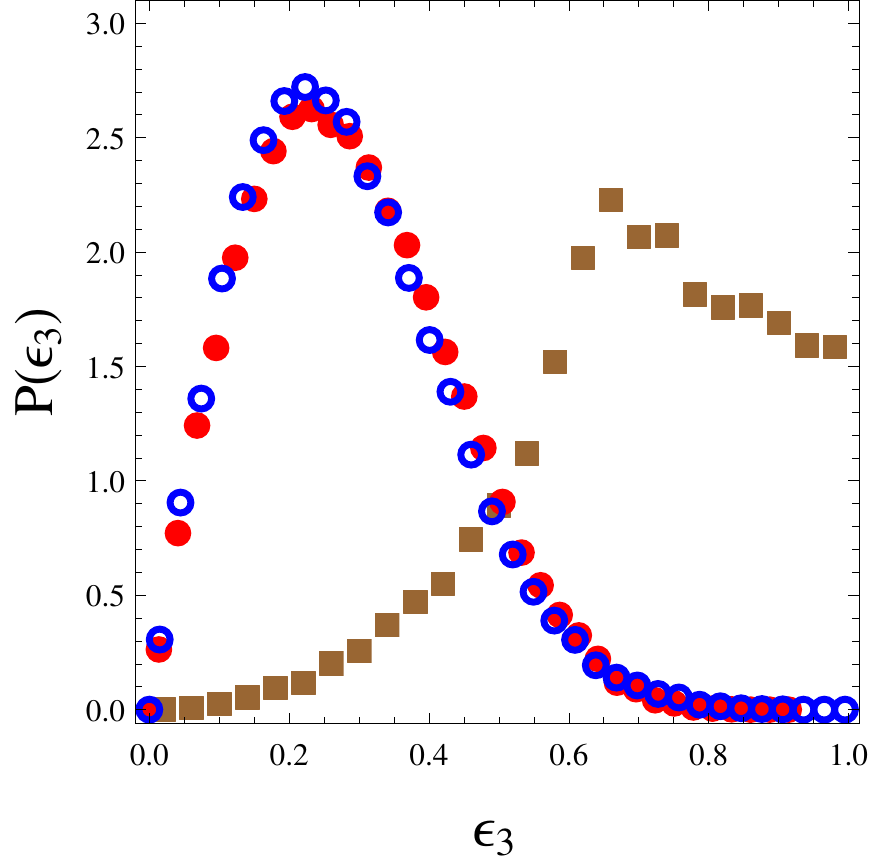}
\caption{Same as Fig.~\ref{fig:eps2} but for the triangularity.
\label{fig:eps3}} 
\end{figure}

Configurations that follow from the $^3$He wave function, with the positions of the nucleons fluctuating, only very rarely realize configurations of maximum triangularity
characteristic of the equilateral triangle, where $\epsilon_2=0$ and $\epsilon_3=1$. 
Indeed, we note widely distributed
$\epsilon_2$ and $\epsilon_3$ in Figs.~\ref{fig:eps2} and \ref{fig:eps3}. 
The  $\epsilon_2$ distribution has a pronounced maximum
at $\epsilon_2 = 1$. These configurations correspond to a very elongated isosceles triangle. 
Such configurations also yield $\epsilon_3\simeq 0.6$, a value corresponding to 
the maximum of the triangularity  distribution  in Fig.~\ref{fig:eps3}. 

Of course, in the collision one does not control the orientation of the nucleus,
which is random. In that case the relevant characteristics of the triangle 
are the  eccentricities evaluated for the triangle projected on the
transverse plane, which is then reflected in the fireball eccentricity.
After projection of the $^3$He configurations with random orientations, the distribution of the ellipticity  is even  more peaked
at $\epsilon_2\simeq 1$, and the  distribution of the  triangularity at $\epsilon_3 \simeq 0.6$. Thus the configurations projected on the 
transverse plane are mostly elongated isosceles triangles. The $^3$He nuclei in such configurations, when hitting the large Au nucleus at 
a small impact parameter, generate a fireball with large $\epsilon_2$ and moderate $\epsilon_3$. 

\section{Eccentricities of the fireball \label{sec:ecc}}

The fireball created in the collision of a small $^3$He nucleus with a large Au target inherits largely 
the shape of the smaller projectile, as discussed in the previous Section. Each of the three He nucleons wounds several nucleons in the Au target. 
The result is a concentration of participant nucleons around the positions of the three He nucleons in the transverse plane.
Therefore, the shape of the fireball preserves partly the ellipticity and triangularity of the incoming $^3$He nucleus, but with 
considerable smearing. Our simulations are carried out with GLISSANDO~\cite{Broniowski:2007nz,Rybczynski:2013yba} and for most of the results use the 
simplest wounded nucleon model. 
We use a realistic wounding profile, which results in a larger smearing than for the black disc case
\cite{Rybczynski:2011wv}. We investigate the RHIC energy of $\sqrt{s_{NN}}=200$~GeV,
where the inelastic NN cross section is equal to 42~mb.  
The source density is obtained  by smearing the density at the Monte-Carlo generated  positions of
the wounded nucleons with a Gaussian of width $0.4$~fm, which introduces a further reduction of azimuthal asymmetries. Additional fluctuations in entropy deposition 
at each source (considered in Sec.~\ref{sec:measure})
smear the initial geometry even more.  While triangularity  increases due to  fluctuations, 
at the same time the imprint of the geometric
triangularity from the deformed $^3$He configuration is washed out to a large degree.

\begin{figure}[tb]
\includegraphics[angle=0,width=0.45 \textwidth]{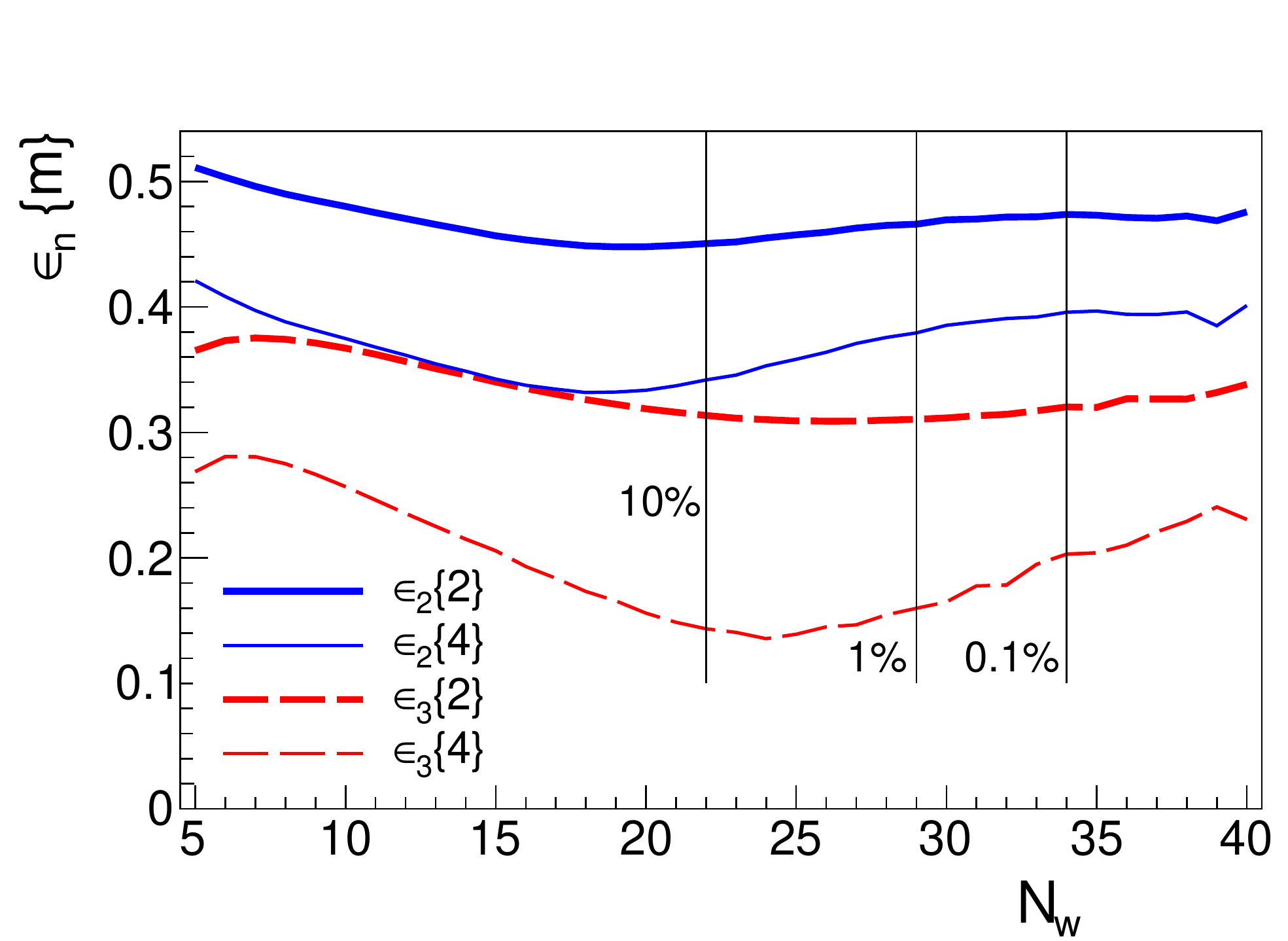}
\caption{Cumulant moments $\epsilon_{n}\{m\}$ from the wounded nucleon model for the 
fireball created in $^{3}$He-$^{197}$Au collisions.
\label{fig:ecc}} 
\end{figure}

The cumulant moments of ellipticity, $\epsilon_2\{2\}$ and $\epsilon_2\{4\}$, are
very large (Fig. \ref{fig:ecc}). Moreover, for centralities below 10\% they do not decrease with the increasing number of wounded nucleons,
which signals a significant contribution of the intrinsic geometric deformation. 
This observation is consistent with the characteristics of $^3$He configurations (Sect. \ref{sec:he3}). 
The projectile $^3$He has a dominant quadrupole deformation, whereas its triangular deformation is significantly smaller.

A similar trend is visible in the dependence of the cumulant moments of $\epsilon_3$ on $N_{w}$, i.e., 
they also do not decrease for the most central events. In fact, the behavior is non-monotonous, 
especially strong for $\epsilon_3\{4\}$. The change in the trend reflects switching from fluctuation-driven triangularity  at smaller $N_{w}$ to 
domination of the intrinsic geometry deformation for the most central events.
The modification of the trend in the centrality dependence of $\epsilon_3$ is probably not strong enough to imply a noticeable 
signature in the centrality dependence of the triangular flow $v_3$. This is because the hydrodynamic response  increases with
the multiplicity of the event
and its effect in small systems depends on details of the hydrodynamic evolution~\cite{Bozek:2013uha}.  
 
In Figs. \ref{fig:eps2} and \ref{fig:eps3} we also show the distributions of eccentricities
of the fireball at $N_{w}=22$ and $N_{w}=34$, corresponding to centralities $10$\% and $0.1$\%, respectively.
We notice that the fireball eccentricities are significantly smaller than the eccentricities of the $^3$He configurations. 
As discussed above, it is due to a random orientation of the incoming $^3$He nucleus, 
and smearing of initial density with the Gaussians centered at the positions 
of the wounded nucleons. Triggering on the most central events increases  
the ellipticity of the fireball, but has a small effect for the average triangularity. Even triggering on ultra-central events ($c < 0.1$\%) is
not enough to provide a direct experimental signature of the geometric triangularity in the system, as the increase of the average triangular flow is not  
much stronger than  the expected increase of $v_3$ from the stronger hydrodynamic response in the very central collisions.

\begin{figure}[tb]
\includegraphics[angle=0,width=0.45 \textwidth]{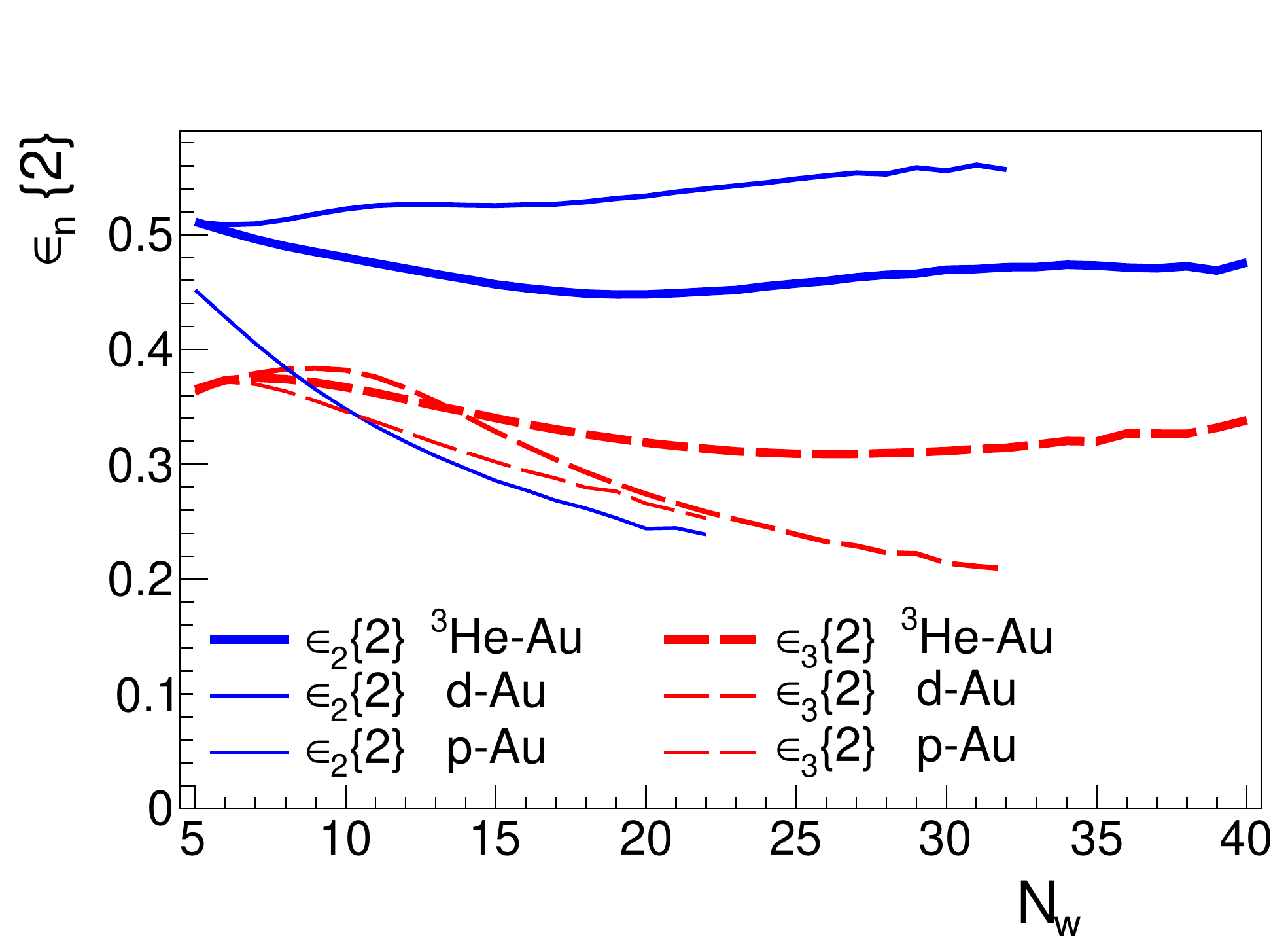}
\caption{Cumulant moments $\epsilon_{n}\{2\}$ from the wounded nucleon model for the fireballs created in various reactions at $200$~GeV.
\label{fig:Hedp}} 
\end{figure}

A comparison of eccentricities for different collision systems: p-Au, d-Au, and $^3$He-Au, exhibits differences that signal different origin of the fireball eccentricities (Fig. \ref{fig:Hedp}). 
The moment $\epsilon_2\{2\}$ is large for d-Au and $^3$He-Au collisions, reflecting the large elliptic deformation of the projectile nucleus. 
The triangularity in p-Au and d-Au collisions originates from fluctuations only, thus  decreases for the central events, while an 
opposite behavior for the geometry-driven triangularity in the $^3$He-Au case is observed. 
Therefore, the comparison of the triangular flow $v_3$ in p-Au and d-Au reactions to the $^3$He-Au case might display the geometric triangularity in the latter. 
However, the argument may be difficult to apply in practice. 
First, the same number on participants corresponds to very different centralities in all the three systems. Second, the hydrodynamic response depends not only on the multiplicity in the system, 
but also on its size, hence relations between eccentricities cannot be compared directly to analogous relation between flow in different systems.

\section{Predictions for measurable quantities \label{sec:measure}}

As stated in Sec.~\ref{sec:method}, a simple way to assess the properties of
the collective flow without hefty hydrodynamic simulations is to consider the
ratios of cumulant moments of the flow coefficients (\ref{eq:ratios}). 
Importantly, these ratios provide simple signatures of the appearance of intrinsic geometry, or just the fluctuation driven flow asymmetry.

\begin{figure}
\includegraphics[angle=0,width=0.45 \textwidth]{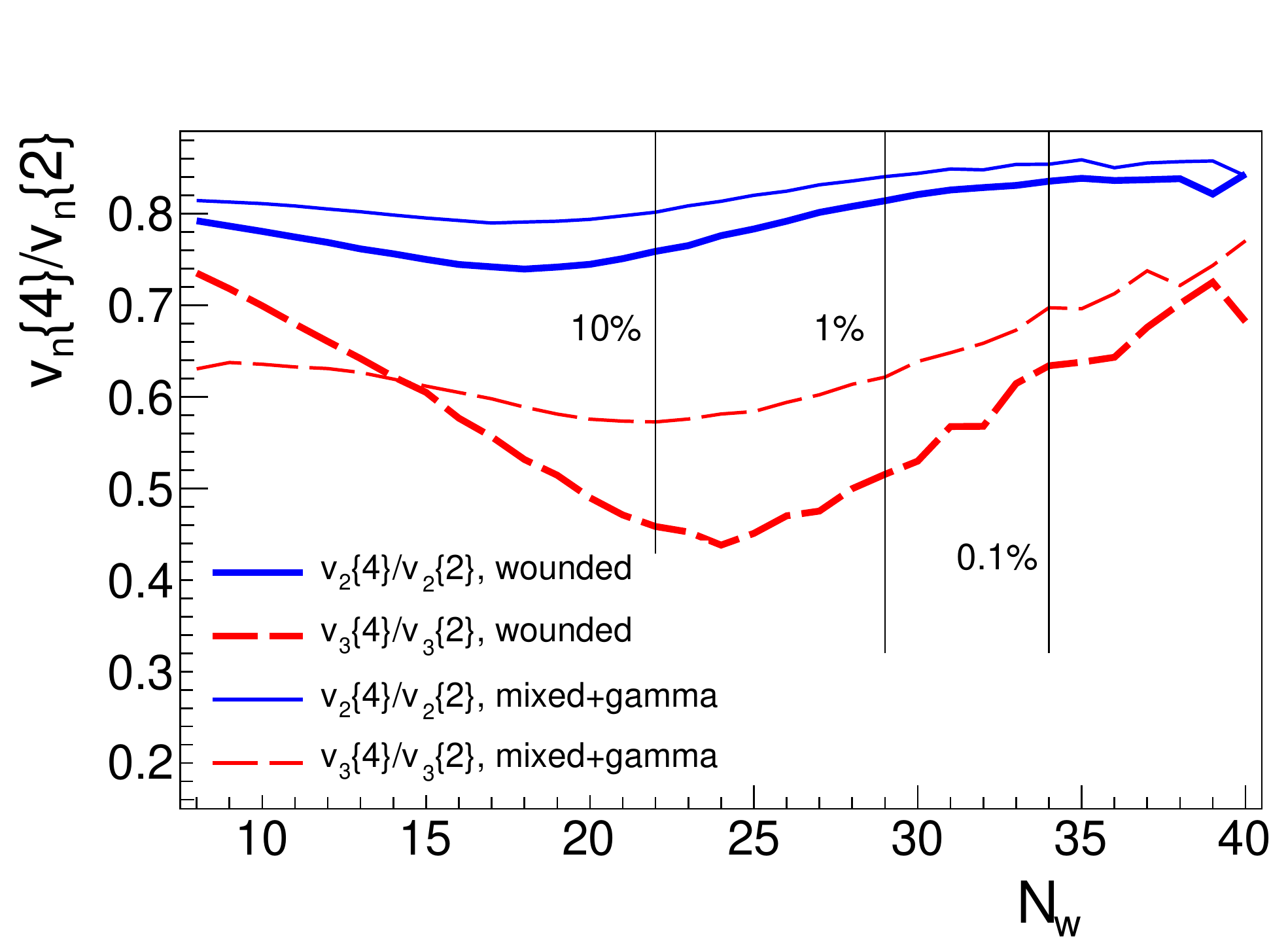} 
\caption{Ratios $v_n\{4\}/v_n\{2\}$, $n=2,3$, for $^{3}$He-$^{197}$Au collisions. The fireball is calculated in two 
version of the Glauber model: the usual wounded nucleon model, and a model with an admixture of binary collisions and with
fluctuations in the  entropy deposition for each participant nucleon (mixed+gamma).
\label{fig:ratios}} 
\end{figure} 

The ratio of the flow coefficients $v_3\{4\}/v_3\{2\}$ is non-monotonous function of $N_w$ (Fig. \ref{fig:ratios}). 
For small $N_{w}$ it decreases as expected from a fluctuation mechanism of source shape deformation. At around $N_{w}=22$ the trend is reversed, signaling the dominance of the
geometric triangular deformation. The change in the trend is due to two reasons. First, by triggering on high-$N_w$ events the orientations of the incoming $^3$He projectile 
become somewhat more deformed. Second, the fluctuations of $v_3$ decrease as the number of participant nucleons increases, and the ratio $v_3\{4\}/v_3\{2\}$ increases towards $1$.
We note that a similar change in the trend of the dependence on $N_w$ is visible for the ratio $v_2\{4\}/v_2\{2\}$. 

The balance between the geometry and fluctuations depends on the model of the initial fireball formation. 
One source of additional fluctuations comes from fluctuations in the entropy deposition
from each participant. Here we use a model with a gamma distribution for the entropy distribution~\cite{Rybczynski:2013yba,Bozek:2013uha} 
superimposed over the distribution of participants.  On the other hand, the admixture of binary collisions in the fireball  
makes the geometric deformation stronger, as the concentration of binary collisions (located in the mean location of the two colliding nucleons) 
follows closer the shape of the $^3$He projectile than the distribution of 
the wounded nucleons from the Au nucleus. In Fig.~\ref{fig:ratios} we show the results of the Glauber model with an admixture
of binary collisions and fluctuation of the deposited entropy. In all variants of the calculations we find that the proposed signature of the geometric flow, 
namely the change in the trend for the ratio $v_2\{4\}/v_2\{2\}$, is still present. 

The described change in the trend for the ratio $v_3\{4\}/v_3\{2\}$ as a function of centrality is the main result of this Letter. 
The centralities where the minimum of the ratio occurs   ($\simeq 10$\%)  
are easily accessible in experimental analysis. Experimentally,
the ratio $v_3\{4\}/v_3\{2\}$ for the centrality bin 5-10\% should be compared to the one in ultra-central events, 0-$0.1$\%, and in 
semi-peripheral events, e.g., 20-40\%., to search for a non-monotonic dependence on centrality.


As the sensitivity of the results shown in Fig.~\ref{fig:ratios} on the fireball formation model is significant, 
precise measurements of this quantity may be used to discriminate between these models. 

Finally, we remark that the results for the $^3$H collisions with the configurations of Ref.\cite{Carlson:1997qn} are 
indistinguishable from the $^3$He case presented in this work.

\section{Conclusion}

The $^3$He-Au collisions form a system where the intrinsic triangular deformation could lead to a large triangular flow. 
Hydrodynamic simulations predict a triangular flow of emitted particles~\cite{Nagle:2013lja}, but the
contribution to the flow from geometry and fluctuations in the initial state cannot be easily separated, since
the relatively small number of participant nucleons gives large fluctuations of the fireball shape.
The  small effect of the intrinsic   triangular deformation of 
the $^3$He projectile on flow  signatures can be traced to a number of reasons:
1)~The most probable three nucleon configurations in $^3$He wave-function have the shape of an elongated triangle with
 $\epsilon_3\simeq 0.6$ and $\epsilon_2 \simeq 1$. 2)~In the collisions, the fireball is determined not by the wave-function configuration, 
but its protection on the transverse plane. The three nucleon configurations projected on the transverse plane 
are even more dominated by configurations with $\epsilon_3\simeq 0.6$ and $\epsilon_2 \simeq 1$.
3)~As a result the  fireball created in a $^3$He-Au collision has most often a very large ellipticity $\epsilon_2$  and 
a smaller triangularity $\epsilon_3$. 4)~Triggering on central events  does not change the average triangularity $\epsilon_3$ significantly.

In that situation, we  propose to look at the ratio $v_3\{4\}/v_3\{2\}$ as a function of centrality, or $N_w$. For $^3$He-Au collisions this ratio has a 
non-monotonic behavior, with a minimum at centrality $10$\%. For collisions with a small number of wounded nucleons, fluctuations dominate the triangularity 
and the ratio $v_3\{4\}/v_3\{2\}$ decreases with increasing $N_{w}$, while for the most central events the geometric deformation dominates, 
and the ratio increases. The main reason to look at  the centrality dependence of the ratio instead of the centrality dependence of $v_3\{2\}$ or $v_3\{4\}$ is that  
a large part of the centrality dependence of $v_3\{m\}$ comes from the change of  hydrodynamic response coefficient with centrality. Moreover, 
the centrality dependence of the hydrodynamic response in small systems is not very well constrained in the models. The proposed signature $v_n\{4\}/v_n\{2\}$ can be 
straightforwardly investigated in experiments with $^3$He-Au collisions at RHIC or, more generally, when looking for elliptic or triangular flow driven by the projectile geometry in d-Au, $^9$Be-Au, or $^{12}$C-Au collisions.

\bigskip

Supported by National Science Centre, Grant No.
DEC-2012/05/B/ST2/02528, DEC-2012/06/A/ST2/00390 and by PL-Grid Infrastructure.

\bigskip

\bibliography{hep,hydr}

\end{document}